\documentclass{article}
\usepackage{amsmath,amssymb,graphicx,enumerate,bbm,theorem,ifthen,pst-all,placeins}

\newcommand{\lp}[1]{\ifthenelse{\equal{#1}{0}}{(}{}\ifthenelse{\equal{#1}{1}}{\bigl(}{}\ifthenelse{\equal{#1}{2}}{\Bigl(}{}\ifthenelse{\equal{#1}{3}}{\biggl(}{}\ifthenelse{\equal{#1}{4}}{\Biggl(}{}\ifthenelse{\equal{#1}{5}}{\Biggl(}{}}
\newcommand{\rp}[1]{\ifthenelse{\equal{#1}{0}}{)}{}\ifthenelse{\equal{#1}{1}}{\bigr)}{}\ifthenelse{\equal{#1}{2}}{\Bigr)}{}\ifthenelse{\equal{#1}{3}}{\biggr)}{}\ifthenelse{\equal{#1}{4}}{\Biggr)}{}\ifthenelse{\equal{#1}{5}}{\Biggr)}{}}
\newcommand{\lbc}[1]{\ifthenelse{\equal{#1}{0}}{\{}{}\ifthenelse{\equal{#1}{1}}{\bigl\{}{}\ifthenelse{\equal{#1}{2}}{\Bigl\{}{}\ifthenelse{\equal{#1}{3}}{\biggl\{}{}\ifthenelse{\equal{#1}{4}}{\Biggl\{}{}\ifthenelse{\equal{#1}{5}}{\Biggl\{}{}}
\newcommand{\rbc}[1]{\ifthenelse{\equal{#1}{0}}{\}}{}\ifthenelse{\equal{#1}{1}}{\bigr\}}{}\ifthenelse{\equal{#1}{2}}{\Bigr\}}{}\ifthenelse{\equal{#1}{3}}{\biggr\}}{}\ifthenelse{\equal{#1}{4}}{\Biggr\}}{}\ifthenelse{\equal{#1}{5}}{\Biggr\}}{}}
\newcommand{\lba}[1]{\ifthenelse{\equal{#1}{0}}{\langle}{}\ifthenelse{\equal{#1}{1}}{\bigl\langle}{}\ifthenelse{\equal{#1}{2}}{\Bigl\langle}{}\ifthenelse{\equal{#1}{3}}{\biggl\langle}{}\ifthenelse{\equal{#1}{4}}{\Biggl\langle}{}\ifthenelse{\equal{#1}{5}}{\Biggl\langle}{}}
\newcommand{\rba}[1]{\ifthenelse{\equal{#1}{0}}{\rangle}{}\ifthenelse{\equal{#1}{1}}{\bigr\rangle}{}\ifthenelse{\equal{#1}{2}}{\Bigr\rangle}{}\ifthenelse{\equal{#1}{3}}{\biggr\rangle}{}\ifthenelse{\equal{#1}{4}}{\Biggr\rangle}{}\ifthenelse{\equal{#1}{5}}{\Biggr\rangle}{}}
\newcommand{\ve}[1]{\ifthenelse{\equal{#1}{0}}{|}{}\ifthenelse{\equal{#1}{1}}{\big|}{}\ifthenelse{\equal{#1}{2}}{\Big|}{}\ifthenelse{\equal{#1}{3}}{\bigg|}{}\ifthenelse{\equal{#1}{4}}{\Bigg|}{}\ifthenelse{\equal{#1}{5}}{\Bigg|}{}}
\newcommand{\lb}[1]{\ifthenelse{\equal{#1}{0}}{[}{}\ifthenelse{\equal{#1}{1}}{\bigl[}{}\ifthenelse{\equal{#1}{2}}{\Bigl[}{}\ifthenelse{\equal{#1}{3}}{\biggl[}{}\ifthenelse{\equal{#1}{4}}{\Biggl[}{}\ifthenelse{\equal{#1}{5}}{\Biggl[}{}}
\newcommand{\rb}[1]{\ifthenelse{\equal{#1}{0}}{]}{}\ifthenelse{\equal{#1}{1}}{\bigr]}{}\ifthenelse{\equal{#1}{2}}{\Bigr]}{}\ifthenelse{\equal{#1}{3}}{\biggr]}{}\ifthenelse{\equal{#1}{4}}{\Biggr]}{}\ifthenelse{\equal{#1}{5}}{\Biggr]}{}}
\newcommand{\srp}[3]{\ifthenelse{\equal{#1}{0}}{)^{#2}_{#3}}{}\ifthenelse{\equal{#1}{1}}{\bigr)^{#2}_{#3}}{}\ifthenelse{\equal{#1}{2}}{\Bigr)^{#2}_{#3}}{}
\ifthenelse{\equal{#1}{3}}{\biggr)^{#2}_{#3}}{}\ifthenelse{\equal{#1}{4}}{\Biggr)^{#2}_{#3}}{}\ifthenelse{\equal{#1}{5}}{\Biggr)^{#2}_{#3}}{}}
\newcommand{\srb}[3]{\ifthenelse{\equal{#1}{0}}{]^{#2}_{#3}}{}\ifthenelse{\equal{#1}{1}}{\bigr]^{#2}_{#3}}{}\ifthenelse{\equal{#1}{2}}{\Bigr]^{#2}_{#3}}{}
\ifthenelse{\equal{#1}{3}}{\biggr]^{#2}_{#3}}{}\ifthenelse{\equal{#1}{4}}{\Biggr]^{#2}_{#3}}{}\ifthenelse{\equal{#1}{5}}{\Biggr]^{#2}_{#3}}{}}
\newcommand{\srbc}[3]{\ifthenelse{\equal{#1}{0}}{{\}}^{#2}_{#3}}{}\ifthenelse{\equal{#1}{1}}{\bigr{\}}^{#2}_{#3}}{}\ifthenelse{\equal{#1}{2}}{\Bigr{\}}^{#2}_{#3}}{}
\ifthenelse{\equal{#1}{3}}{\biggr{\}}^{#2}_{#3}}{}\ifthenelse{\equal{#1}{4}}{\Biggr{\}}^{#2}_{#3}}{}\ifthenelse{\equal{#1}{5}}{\Biggr{\}}^{#2}_{#3}}{}}
\newcommand{\srba}[3]{\ifthenelse{\equal{#1}{0}}{\rangle^{#2}_{#3}}{}\ifthenelse{\equal{#1}{1}}{\bigr\rangle^{#2}_{#3}}{}\ifthenelse{\equal{#1}{2}}{\Bigr\rangle^{#2}_{#3}}{}
\ifthenelse{\equal{#1}{3}}{\biggr\rangle^{#2}_{#3}}{}\ifthenelse{\equal{#1}{4}}{\Biggr\rangle^{#2}_{#3}}{}\ifthenelse{\equal{#1}{5}}{\Biggr\rangle^{#2}_{#3}}{}}
\newcommand{\sve}[3]{\ifthenelse{\equal{#1}{0}}{|^{#2}_{#3}}{}\ifthenelse{\equal{#1}{1}}{\bigr|^{#2}_{#3}}{}\ifthenelse{\equal{#1}{2}}{\Bigr|^{#2}_{#3}}{}
\ifthenelse{\equal{#1}{3}}{\biggr|^{#2}_{#3}}{}\ifthenelse{\equal{#1}{4}}{\Biggr|^{#2}_{#3}}{}\ifthenelse{\equal{#1}{5}}{\Biggr|^{#2}_{#3}}{}}
\newcommand{\im}{\mathcal{I}m}
\newcommand{\re}{\mathcal{R}e}
\newcommand{\sh}{\ensuremath{\tmop{sh}}}
\newcommand{\ch}{\ensuremath{\tmop{ch}}}

\newcommand{\cylindreb}[1]{\ensuremath{\tilde{S}_{#1}}}

\newcommand{\disqueb}[1]{\ensuremath{D_{#1}}}
\newcommand{\demidisqueb}[1]{\ensuremath{D_{#1}^+}}
\newcommand{\demiplanb}{\ensuremath{\mathbbm{C}^+}}
\newcommand{\nevb}[1]{\ensuremath{\tilde{D}_{#1}}}

\newcommand{\disque}[1]{{\disqueb{#1}}}
\newcommand{\demidisque}[1]{{\demidisqueb{#1}}}
\newcommand{\demiplan}[1]{{\demiplanb}#1}
\newcommand{\nev}[1]{{\nevb{#1}}}
\newcommand{\cylindre}[1]{{\cylindreb{#1}}}

\newcommand{\AMSclass}[1]{{\textbf{A.M.S. subject classification:} #1}}
\newcommand{\assign}{:=}
\newcommand{\keywords}[1]{{\textbf{Keywords:} #1}}
\newcommand{\nin}{\not\in}
\newcommand{\nonesep}{}
\newcommand{\tmmathbf}[1]{\ensuremath{\boldsymbol{#1}}}
\newcommand{\tmop}[1]{\ensuremath{\operatorname{#1}}}
\newcommand{\tmscript}[1]{\text{\scriptsize{$#1$}}}
\newcommand{\tmtextbf}[1]{{\bfseries{#1}}}
\newcommand{\tmtextup}[1]{{\upshape{#1}}}
\newcommand{\um}{-}
\newenvironment{enumeratenumeric}{\begin{enumerate}[1.] }{\end{enumerate}}
\newenvironment{itemizeminus}{\begin{itemize} }{\end{itemize}}
\newenvironment{proof}{\noindent\textbf{Proof\ }}{\hspace*{\fill}$\Box$\medskip}
\numberwithin{equation}{section}  
\numberwithin{figure}{section}  
\newtheorem{theorem}{Theorem}[section]
\newtheorem{lemma}[theorem]{Lemma}
\newtheorem{proposition}[theorem]{Proposition}
\newtheorem{corollary}[theorem]{Corollary}
\newtheorem{definition}[theorem]{Definition}
\newtheorem{remark}[theorem]{Remark}

\newtheorem{example}[theorem]{Example}

\begin{document}

\title{Some remarks on the complex heat kernel on $\mathbbm{C}^{\nu}$ in the
scalar potential case\thanks{This paper has been written using the GNU TEXMACS scientific text editor.}
\thanks{\keywords{heat kernel, asymptotic expansion,
Wigner-Kirkwood expansion, complex variables, Borel summation};
\AMSclass{30E15, 32W30, 35K05, 35K08, 35C20}}}\author{Thierry
Harg\'e}\date{January 25, 2013}\maketitle

\begin{abstract}
  In previous works, we used a so-called deformation formula in order to
  study, in particular, the Borel summability of the heat kernel of some
  operators. A goal of this paper is to collect miscellaneous remarks related
  to these works. Here the complex setting plays an important role. Moreover,
  the deformation formula provides a solution of the heat equation in
  ``unusual'' cases. We also give a uniqueness statement concerning these
  cases.
\end{abstract}

\section{Introduction}

In previous works [Ha4, Ha5], we used a so-called deformation formula in order
to study the Borel summability of the heat kernel, $p (t, x, y)$, associated
to a $\nu$-dimensional partial differential operator ($\nu \in
\mathbbm{N}^{\ast})$. This formula is extended to the non-autonomous case
[Ha6]. The natural setting for this formula is a complex one ($t \in
\mathbbm{C}, \re t \geqslant 0$ and $x, y \in \mathbbm{C}^{\nu}$). For
instance this formula is valid for operators as
\begin{equation}
  \label{intromonica2} P \assign \partial_x^2 + \lambda x^2 + c (x) \assign
  \partial_{x_1}^2 + \cdots + \partial_{x_{\nu}}^2 + \lambda (x_1^2 + \cdots +
  x_{\nu}^2) + c (x_1, \ldots, x_{\nu})
\end{equation}
where $\lambda \in \mathbbm{R}$ and the function $c$ is the Fourier transform
of a suitable Borel measure. The aim of this paper is threefold:
\begin{enumeratenumeric}
  \item Defining in a unique way the heat kernel associated to the operator
  $P$ is a well known procedure if $\lambda \leqslant 0$. If $\lambda > 0$,
  one can use the commutator theorem [Re-Si] for instance (see Remark
  \ref{unicityolga6.65}). However, we look for a statement adapted to a full
  complex setting and covering the non-autonomous case: $P = P_0 + c$ where
  $P_0$ is defined by (\ref{unicityolga1}). Here is the purpose of Proposition
  \ref{unicityolga6}. The statement and the proof of this proposition are
  standard. We assume that the coefficients defining $P_0$ satisfy a
  reality-preserving property; see (\ref{unicityolga1.5}). This implies that
  the operator $P_0 |_{i\mathbbm{R}}$ is symmetric with respect to the $L^2$
  inner product. The unicity is therefore a consequence of the conservation of
  the $L^2$-norm for the solutions of the time dependent Schr\"odinger
  equation associated to $P_0$. Let us make the following remarks.
  \begin{itemizeminus}
    \item In our setting, the Schr\"odinger kernel is viewed as the boundary
    value of the heat kernel for imaginary{\footnote{With our notation, a
    physical interpretation of the parameter $t$ is the reciprocal of the
    temperature which is a macroscopic variable. \ We denote by $\tmmathbf{t}$
    the ``physical'' time.}} values of $t$.
    
    \item Our goal is not to solve the heat equation but to study the heat
    kernel as a function defined on some subset of $\mathbbm{C}_{\sqrt{\cdot}}
    \times \mathbbm{C}^{2 \nu}$ ($\mathbbm{C}_{\sqrt{\cdot}}$ denotes the
    Riemann{\footnote{A ramification occurs in our statements if the dimension
    $\nu$ is odd: this is only due to the existence of the factor $t^{- \nu /
    2}$ in the expression of the free heat kernel.}} surface of the square
    root function).
  \end{itemizeminus}
  About the existence problem, we use [Ha4] and [Ha6].
  
  \item Considering a complex setting allows one to make some remarks.
  
  We reformulate Proposition \ref{unicityolga6} by using the analytic dilation
  given by
  \begin{equation}
    \label{intromonica4} (t, x, y) \mapsto (e^{i \epsilon} t, e^{i \epsilon /
    2} x, e^{i \epsilon / 2} y) \text{ , \ } \epsilon \in \mathbbm{R}/ 4 \pi
    \mathbbm{Z}
  \end{equation}
  (see Corollary \ref{rotatedanna4}). As a consequence, we see that the
  deformation formula provides a solution for the Schr\"odinger equation
  \[ i^{- 1} \partial_{\tmmathbf{t}} p = \lp{1} \partial_x^2 + \lambda x^2 +
     c (x) \rp{1} p \text{ , \ } p|_{\tmmathbf{t} = 0} = \delta_{x = y} \text{
     , \ } ( \tmmathbf{t} \in \mathbbm{R}, | \tmmathbf{t} | \ll 1, x, y \in
     \mathbbm{R}^{\nu}) \]
  with fast growing potentials such as, for instance, $c (x) = e^{x^2}$ (see
  Corollary \ref{rotatedanna6.1} and Remark \ref{rotatedanna8.5}). We do not
  state uniqueness in this case (this can be done by keeping the complex point
  of view for the space variables). This equation, in the free case ($\lambda
  = 0$), was considered by Kuna, Streit and Westerkamp [K-S-W]. In this paper,
  the authors build a Feynman integral dealing with such potentials. A version
  of the deformation formula can also be found there [K-S-W, Remark 19].
  
  Then we give a simple assumption on the potential $c$ providing the
  existence of the heat kernel on a conical neighboorhood of $\mathbbm{R}^+$
  but with an aperture larger than $\pi / 2$ (see Proposition
  \ref{rotatedanna9}).
  
  \item We consider the Borel summability of the small time expansion of the
  conjugate of the heat kernel in the free case ($\lambda = 0$): we
  reformulate a statement given in [Ha4] by using the analytic dilation given
  by (\ref{intromonica4}). As in Proposition \ref{rotatedanna9}, we consider
  two cases. We consider a simple class of potentials for which
  Borel-Nevalinna summability (see Section \ref{borelsummation} for the
  definition) holds in an arbitrary direction. Then we give assumptions on the
  potential $c$ implying Borel-Watson summability of this small time expansion
  instead of Borel-Nevalinna summability as in [Ha4]. Borel-Nevalinna
  summability uses that $c$ is the Fourier transform of some Borel measure
  $\mu$ defined on $\mathbbm{R}^{\nu}$ with a suitable convergence assumption.
  Borel-Watson summability uses that $c$ is the Fourier transform of an
  analytic function defined on a conical neighbourhood of $\mathbbm{R}^{\nu}$,
  with a similar convergence assumption. Let us remark that Borel-Watson
  summability of a series is a central tool when the critical time is a
  non-trivial power of the variable [Bals], [Ma-Ra].
\end{enumeratenumeric}
We assume in this paper that the potential $c$ is $\mathbbm{C}$-valued. Our
statements also hold if this potential is matrix-valued as in [Ha4] and [Ha6].

\section{\label{notations}Notation}

For $z \in \mathbbm{C}$, we denote $\sh z \assign \frac{1}{2} (e^z - e^{-
z})$, $\ch z \assign \frac{1}{2} (e^z + e^{- z})$. Let $\nu \geqslant 1$. For
$\lambda, \mu \in \mathbbm{C}^{\nu}$, we denote $\lambda \cdot \mu \assign
\lambda_1 \mu_1 + \cdots + \lambda_{\nu} \mu_{\nu}$, $\lambda^2 \assign
\lambda \cdot \lambda$. These notations are extended to operators as
$\partial_x \assign (\partial_{x_1}, \ldots, \partial_{x_{\nu}})$. We also
denote \ $\bar{\lambda} \assign ( \bar{\lambda}_1, \ldots,
\bar{\lambda}_{\nu})$, $| \lambda | \assign (\lambda \cdot \bar{\lambda})^{1 /
2}$ (if $\lambda \in \mathbbm{R}^{\nu}$, $| \lambda | = \sqrt{\lambda^2}$). We
denote by $\tmmathbf{\pi}$ the canonical projection from $\mathbbm{R}/ 4 \pi
\mathbbm{Z}$ onto $\mathbbm{R}/ 2 \pi \mathbbm{Z}$. For $\theta \in
\mathbbm{R}/ 4 \pi \mathbbm{Z}$, we denote by $e^{i \theta}$ the element of
$\mathbbm{C}_{\sqrt{\cdot}}$, the Riemann surface of the square root function,
with argument $\theta$ and modulus $1$. Then $\mathbbm{C}_{\sqrt{\cdot}}
\assign \{z = r e^{i \theta} |r > 0, \theta \in \mathbbm{R}/ 4 \pi
\mathbbm{Z}\}$. For $z = r e^{i \theta} \in \mathbbm{C}_{\sqrt{\cdot}}$, we
denote $z^{1 / 2} \assign r^{1 / 2} e^{i \theta / 2}$. If $z \in \mathbbm{C}$,
we also denote by $z^{1 / 2}$ the square root of $z$ which is defined up to a
sign. Let $m \geqslant 1$ and $\theta \in \mathbbm{R}/ 2 \pi \mathbbm{Z}$. For
every subset $A$ of $\mathbbm{C}^m$, we denote $e^{i \theta} A \assign \{e^{i
\theta} z |z \in A\}$. We also use this notation if $\theta \in \mathbbm{R}/ 4
\pi \mathbbm{Z}$ and $A \subset \mathbbm{C}_{\sqrt{\cdot}}$. If $\alpha \in
\mathbbm{R}/ 2 k \pi \mathbbm{Z}$ ($k \in \mathbbm{N}^{\ast})$ and $r \in
\mathbbm{R}^+$, we denote by $] \alpha - r, \alpha + r [_{2 k \pi}$ the open
interval{\footnote{$r > k \pi \Rightarrow \text{$] \alpha - r, \alpha + r [_{2
k \pi} =\mathbbm{R}/ 2 k \pi \mathbbm{Z}$}$.}} of $\mathbbm{R}/ 2 k \pi
\mathbbm{Z}$ with end points $\alpha - r$ and $\alpha + r$. We denote
\[ \demiplan{\assign \{z \in \mathbbm{C} | \re (z) > 0\}} \text{ , \ }
   \overline{\mathbbm{C}^+} \assign \{z \in \mathbbm{C} | \re (z) \geqslant
   0\} \]
and, if $T > 0$,
\[ \disque{T} \assign \{z \in \mathbbm{C} | |z| < T\} \text{ , \ }
   \demidisque{T} \assign \disque{T} \cap \demiplan{} \text{ , \ } \bar{D}^+_T
   \assign \disque{T} \cap \overline{\demiplan{}} . \]
Let $\mathfrak{B}$ denote the collection of all Borel sets on $\mathbbm{R}^m$.
An $\mathbbm{C}$-valued measure $\mu$ on $\mathbbm{R}^m$ is an
$\mathbbm{C}$-valued function on $\mathfrak{B}$ satisfying the classical
countable additivity property (see [Ru]). We denote by $| \mu |$ the positive
measure defined \ by
\[ | \mu | (E) = \sup \sum_{j = 1}^{\infty} | \mu (E_j) | (E \in
   \mathfrak{B}), \]
the supremum being taken over all partitions $\{E_j \}$ of $E$. In particular,
$| \mu | (\mathbbm{R}^m) < \infty$.

We denote by $\mathcal{D}(e^{i \theta} \mathbbm{R}^m)$ the space of smooth
functions with compact support defined on $e^{i \theta} \mathbbm{R}^m$. If
$\Omega$ is an open domain in $\mathbbm{C}^m$, we denote by $\mathcal{A}
(\Omega)$ the space of $\mathbbm{C}$-valued analytic functions on $\Omega$.
These spaces are equipped with their standard Frechet structure (the
semi-norms are indexed by compact sets and eventually differentiation order).
If $U = e^{i \theta}] - T, T [$ or $e^{i \theta} \bar{D}^+_T$ or
$\overline{\demiplan{}}$ and $\mathcal{F}$ is a Frechet space,
$\mathcal{C}^{\infty} \lp{1} U, \mathcal{F} \rp{1}$ denotes the standard
Frechet space of smooth functions defined on $U$ with values in $\mathcal{F}$.
For instance, if the topology of $\mathcal{F}$ is defined by a family of
semi-norms $(| \cdot |_j)_{j \in J}$, the Frechet structure of
$\mathcal{C}^{\infty} \lp{1} U, \mathcal{F} \rp{1}$ is defined by the
semi-norms $| \cdot |_{\alpha, j}$ ($\text{$| f|_{\alpha, j}$} = \sup_{t \in
U} | \partial_t^{\alpha} f (t) |_j$) if $U = e^{i \theta}] - T, T [$ or $e^{i
\theta} \bar{D}^+_T$. In the case $U = \overline{\demiplan{}}$, suprema
defining the semi-norms are taken over compact sets $\overline{\demiplan{}}
\cap D (0, R)$ where $R > 0$. We now define global spaces (with respect to the
space variable). We denote by $\mathcal{S}(\mathbbm{R}^m)$ the space of
Schwartz functions:
\[ f \in \mathcal{S}(\mathbbm{R}^m) \Leftrightarrow \forall (\alpha, \beta)
   \in \mathbbm{N}^m \times \mathbbm{N}^m, \sup_{z \in \mathbbm{R}^m}
   |z^{\alpha} \partial_z^{\beta} f (z) | < \infty . \]
We consider the following spaces of smooth functions
\begin{itemize}
  \item if $I =] - T, T [$ or $I = i] - T, T [$
  \[ f \in \text{$\mathcal{C}^{\infty}_{b, 1} (I \times \mathbbm{R}^m)$}
     \Leftrightarrow \]
  \[ \forall (\alpha, \beta) \in \mathbbm{N} \times \mathbbm{N}^m, \exists C >
     0, \forall (t, z) \in I \times \mathbbm{R}^m, | \partial_t^{\alpha}
     \partial^{\beta}_z f (t, z) | \leqslant C (1 + |z|)^{\alpha} . \]
  \item if $I =\mathbbm{R}$ or $I = i\mathbbm{R}$
  \[ f \in \text{$\mathcal{C}^{\infty}_b (I \times \mathbbm{R}^m)$}
     \Leftrightarrow \forall (\alpha, \beta) \in \mathbbm{N} \times
     \mathbbm{N}^m, \sup_{(t, z) \in I \times \mathbbm{R}^m} |
     \partial_t^{\alpha} \partial^{\beta}_z f (t, z) | < \infty . \]
\end{itemize}

\section{\label{unicity}The
heat kernel viewed from the positive direction}

\subsection{The setting}

Let $U$ be a complex open neighbourhood of $0 \in \mathbbm{C}$. Let $P_0$ be
the operator acting on $\mathcal{A}(U \times \mathbbm{C}^{\nu})$, \ defined by
\begin{equation}
  \label{unicityolga1} P_0 = A (t) \cdot (\partial_x + B (t) x) \otimes
  (\partial_x + B (t) x) - C (t) \cdot x \otimes x
\end{equation}
where $A$, $B$ and $C$ are $\nu \times \nu$ complex matrix-valued analytic
functions on $U$. We assume that the matrices $A$ and $C$ are symmetric and
that the matrix $A (0)$ is real positive definite. We assume that
\begin{equation}
  \label{unicityolga1.5} \text{the functions } A|_{i\mathbbm{R}}, (i B)
  |_{i\mathbbm{R}} \text{ and } C|_{i\mathbbm{R}} \text{ are real-valued near
  } 0.
\end{equation}
Then the equation
\[ \partial_t u = P_0 u \]
with the boundary condition
\begin{equation}
  \label{unicityolga2} u_{} |_{t = 0^+} =_{} \delta_{x = y}
\end{equation}
admits an explicit solution
\begin{equation}
  \label{unicityolga2.5} p^0 \assign \frac{k (t)}{(4 \pi \Delta t)^{\nu / 2}}
  e^{- \frac{1}{4 t} A^{- 1} (0) \cdot (x - y)^2 + Q_t (x, y)} .
\end{equation}
Here $\Delta \assign \det \lp{1} A_{j, k} (0) \srp{1}{}{1 \leqslant j, k
\leqslant \nu}$, the function $k$ is analytic near $0$ and $Q_t$ denotes a
polynomial of total degree at most 2 in $x, y$ whose coefficients are analytic
near 0. Moreover these coefficients take their values in $i\mathbbm{R}$ if $x,
y \in \mathbbm{R}^{\nu}$ and $t \in i\mathbbm{R}$, $|t|$ small enough (see
[Ha6] and in particular Lemma 3.4, assertion 3). By (\ref{unicityolga2}) we
mean that for every $\varphi \in \mathcal{D}(\mathbbm{R}^{\nu})$, $x \in
\mathbbm{R}^{\nu}$ and $\theta \in [- \pi / 2, \pi / 2]$
\[ \int_{\mathbbm{R}^{\nu}} u (r e^{i \theta}, x, y) \varphi (y) d y
   \longrightarrow_{r \rightarrow 0^+} \varphi (x) . \]
Let us consider two particular cases. Let $\lambda \in \mathbbm{R}$ and let
$\omega \assign 2 (- \lambda)^{1 / 2}$. Let us denote
\[ p^{\tmop{harm}} \assign \left( 4 \pi \frac{\tmop{sh} (\omega t)}{\omega}
   \right)^{- \nu / 2} \exp \lp{2} - \frac{1}{4} \frac{\omega}{\tmop{sh}
   (\omega t)} (\tmop{ch} (\omega t) (x^2 + y^2) - 2 x \cdot y) \rp{2} \]
and
\[ p^{\tmop{free}} = (4 \pi t)^{- \nu / 2} e^{- (x - y)^2 / 4 t} . \]
Then $p^{\tmop{harm}}$ (respectively $p^{\tmop{free}}$) satisfies
\[ \partial_t u = \lp{1} \partial_x^2 + \lambda x^2 \rp{1} u \]
(respectively $\partial_t u = \partial_x^2 u$) with the boundary condition
(\ref{unicityolga2}).

\subsection{An existence and uniqueness statement}

In this section, our aim is to give an existence and uniqueness statement
concerning the equation
\begin{equation}
  \label{unicityolga4} \partial_t u = \lp{1} P_0 + c (t, x) \rp{1} u
\end{equation}
We also use the following definition in this section.

\begin{definition}
  \label{unicityolga5}Let $T_b > 0$. Let $f$ be a measurable
  $\mathbbm{C}$-valued function on $\disque{T_b} \times \mathbbm{R}^{\nu}$,
  analytic with respect to the first variable and let $\mu_{\ast}$ be a
  positive measure on $\mathbbm{R}^{\nu}$ such that for every $R > 0$
  \[ \int_{\mathbbm{R}^{\nu}} \exp (R| \xi |) \sup_{|t| < T_b} |f (t, \xi) |d
     \mu_{\ast} (\xi) < \infty . \]
  We denote by $c$ the function belonging to $\mathcal{A}( \disque{T_b} \times
  \mathbbm{\mathbbm{C}}^{\nu})$ defined by
  \[ c (t, x) = \int \exp (ix \cdot \xi) f (t, \xi) d \mu_{\ast} (\xi) . \]
\end{definition}

\begin{proposition}
  \label{unicityolga6}Let $P_0$ be as above and let $T_b > 0$. There exists $T
  > 0$ such that, for every $f$ and $\mu_{\ast}$ as in Definition
  \ref{unicityolga5}, the following assertions hold.
  \begin{enumerate}
    \item \label{unicityolga6.12}For every $\varphi \in
    \mathcal{S}(\mathbbm{R}^{\nu})$ there exists a unique $\psi \in
    \mathcal{C}^{\infty} \lp{1} i] - T, T [, \mathcal{S}(\mathbbm{R}^{\nu})
    \rp{1}$ such that
    \begin{equation}
      \label{unicityolga6.14} \left\{ \begin{array}{l}
        \partial_t \psi = \lp{1} P_0 + c (t, x) \rp{1} \psi\\
        \\
        \psi |_{t = 0} = \varphi
      \end{array} \right.
    \end{equation}
    and the mapping $\varphi \longmapsto \psi$ is continuous from
    $\mathcal{S}(\mathbbm{R}^{\nu})$ onto $\mathcal{C}^{\infty} \lp{1} i] - T,
    T [, \mathcal{S}(\mathbbm{R}^{\nu}) \rp{1}$.
    
    \item \label{unicityolga6.2}Let $p = p (t, x, y)$ be the kernel of the
    operator $\mathcal{P}_t : \mathcal{S}(\mathbbm{R}^{\nu}) \rightarrow
    \mathcal{S}(\mathbbm{R}^{\nu})$ defined by $\mathcal{P}_t (\varphi) = \psi
    (t, \cdot)$ for $t \in i] - T, T [$. Then $p$ can be uniquely extended as
    a function in $\mathcal{C}^{\infty} \lp{1} \bar{D}^+_T -\{0\},
    \mathcal{A}(\mathbbm{C}^{2 \nu}) \rp{1} \cap \mathcal{A} \lp{1} (
    \demidisque{T} -\{0\}) \times \mathbbm{C}^{2 \nu} \rp{1}$. This function
    satisfies (\ref{unicityolga4}) on $( \bar{D}^+_T -\{0\}) \times
    \mathbbm{C}^{2 \nu}$ and (\ref{unicityolga2}). Moreover
    \[ p = p^0 \times p^{\tmop{conj}} \]
    where $p^{\tmop{conj}} = p^{\tmop{conj}} (t, x, y) \in$
    \[ \mathcal{A}( \demidisque{T} \times \mathbbm{C}^{2 \nu}) \cap
       \mathcal{C}^{\infty} \lp{1} \bar{D}^+_T, \mathcal{A}(\mathbbm{C}^{2
       \nu}) \rp{1} \cap \mathcal{C}^{\infty}_{b, 1} (i] - T, T [\times
       \mathbbm{R}^{2 \nu}) . \]
    \item \label{unicityolga6.4}Let us assume that $P_0 = \partial_x^2$ (free
    case). Then
    \[ p^{\tmop{conj}} \in \mathcal{A}( \demiplan{} \times \mathbbm{C}^{2
       \nu}) \cap \mathcal{C}^{\infty} \lp{1} \overline{\demiplan{}},
       \mathcal{A}(\mathbbm{C}^{2 \nu}) \rp{1} \cap \mathcal{C}^{\infty}_b
       (i\mathbbm{R} \times \mathbbm{R}^{2 \nu}) . \]
  \end{enumerate}
\end{proposition}

\begin{remark}
  The spaces $\mathcal{C}^{\infty} \lp{1} \bar{D}^+_T,
  \mathcal{A}(\mathbbm{C}^{2 \nu}) \rp{1}$ and $\mathcal{A}( \demidisque{T}
  \times \mathbbm{C}^{2 \nu})$ are local in the following meaning: the
  semi-norms are defined by taking suprema over compact sets. The spaces
  $\mathcal{S}(\mathbbm{R}^{\nu})$ and $\mathcal{C}^{\infty}_{b, 1} (i] - T, T
  [\times \mathbbm{R}^{2 \nu})$ are global, which is useful for the uniqueness
  statement. By (\ref{unicityolga2.5})
  \[ p^0 = p^{\tmop{free}} \times p^1 \]
  where $p^1 \in \mathcal{A}(D_T \times \mathbbm{C}^{2 \nu})$ for some $T >
  0$. However one can not replace $p^0$ by $p^{\tmop{free}}$ in the statement
  of Proposition \ref{unicityolga6} since $p^1 \nin \mathcal{C}^{\infty}_{b,
  1} (i] - T, T [\times \mathbbm{R}^{2 \nu})$ in the general case.
\end{remark}

\begin{remark}
  \label{unicityolga6.64}The reality-preserving property
  (\ref{unicityolga1.5}) is useful for the uniqueness statement. This
  assumption implies that the operator $P_0 |_{i\mathbbm{R}}$ is symmetric
  with respect to the $L^2$ inner product.
\end{remark}

\begin{remark}
  \label{unicityolga6.65}Let us assume that $P_0 = \partial_x^2 + \lambda x^2$
  where $\lambda \in \mathbbm{R}$, that the potential $c$ does not depend on
  $t$. Since $c \in L^{\infty} (\mathbbm{R}^{\nu})$, by [Re-Si, Th. X.36], the
  operator $P_0 + c$ is self-adjoint on a domain containing the domain of the
  operator $- \partial_x^2 + x^2$. See also [Bo-Ca-H\"a-Mi]. Therefore one can
  define the Schr\"odinger operator for the operator $P_0$. By [Ha4], its
  kernel admits a (unique) analytic continuation on $( \bar{D}^+_T -\{0\})
  \times \mathbbm{C}^{2 \nu}$ if $T$ is small enough. This yields an
  alternative formulation of Proposition \ref{unicityolga6} in the harmonic
  case.
\end{remark}

\subsection{Proof of Proposition \ref{unicityolga6}}

We need the following lemma.

\begin{lemma}
  \label{unicityolga8}Let $T > 0$ and let $B$ be a $\nu \times \nu$ real
  positive definite symmetric matrix. Let $P_{\tmmathbf{t}}$ be a polynomial
  with respect to $x, y \in \mathbbm{R}^{\nu}$, of degree at most $2$, with
  coefficients belonging to $\mathcal{C}^{\infty} (] - T, T [, \mathbbm{R})$.
  There exists $T_2 > 0$ such that, for every $u \in \mathcal{C}^{\infty}_{b,
  1} (] - T, T [\times \mathbbm{R}^{2 \nu})$ and every $\varphi \in
  \mathcal{S}(\mathbbm{R}^{\nu})$, the function $\tmmathbf{\psi}$ defined by
  \begin{equation}
    \label{unicityolga8.1} \tmmathbf{\psi} ( \tmmathbf{t}, x) \assign
    \int_{\mathbbm{R}^{\nu}} \lp{2} \frac{\det B}{4 \pi i \tmmathbf{t}}
    \srp{2}{\nu / 2}{} e^{- B \cdot (x - y)^2 / 4 i \tmmathbf{t}} e^{i
    \nonesep P_{\tmmathbf{t}} (x, y)} u ( \tmmathbf{t}, x, y) \varphi (y) d y
  \end{equation}
  belongs to $\mathcal{C}^{\infty} \lp{1}] - T_2, T_2 [,
  \mathcal{S}(\mathbbm{R}^{\nu}) \rp{1}$. The mapping $\varphi \longmapsto
  \tmmathbf{\psi}$ is continuous from $\mathcal{S}(\mathbbm{R}^{\nu})$ into
  $\mathcal{C}^{\infty} \lp{1}] - T_2, T_2 [, \mathcal{S}(\mathbbm{R}^{\nu})
  \rp{1}$. Moreover, if $e^{i \nonesep P_0 (y, y)} \times u (0, y, y) = 1$,
  then $\tmmathbf{\psi} |_{\tmmathbf{t} = 0} = \varphi$.
\end{lemma}

Such a result is standard. The proof is given in the Appendix.

Let us prove Proposition \ref{unicityolga6}. The assertion
\ref{unicityolga6.12}, of course, is a well-known statement. However, for the
convenience of the reader and the completeness of the paper, we give its
proof.

\medskip

\tmtextbf{-1-} In view of the uniqueness statement in assertion
\ref{unicityolga6.12}, let us consider $\psi \in \mathcal{C}^{\infty} \lp{1}
i] - T, T [, \mathcal{S}(\mathbbm{R}^{\nu}) \rp{1}$ satisfying
(\ref{unicityolga6.14}) with $\varphi = 0$. Let
\[ E ( \tmmathbf{t}) \assign \int_{\mathbbm{R}^{\nu}} \psi (i \tmmathbf{t}, x)
   \bar{\psi} (i \tmmathbf{t}, x) d x, \]
\[ R ( \tmmathbf{t}) \assign - 2 \int_{\mathbbm{R}^{\nu}} \im \lp{1} c (i
   \tmmathbf{t}, x) \rp{1} \psi (i \tmmathbf{t}, x) \bar{\psi} (i
   \tmmathbf{t}, x) d x. \]
Then by (\ref{unicityolga1.5}), for every $\tmmathbf{t} \in] - T, T [$,
$\partial_{\tmmathbf{t}} E = R ( \tmmathbf{t})$. Since the function $|c|$ is
bounded, there exists $K > 0$ such that $| \partial_{\tmmathbf{t}} E |
\leqslant K E$. Since $\psi \bar{\psi}$ is bounded by $C (1 + |x|)^{- \nu -
1}$ for $( \tmmathbf{t}, x) \in] - T, T [\times \mathbbm{R}^{\nu}$, one gets
$E|_{\tmmathbf{t} = 0} = 0$ by the dominated convergence theorem. Therefore $E
= 0$ and $\psi |_{i] - T, T [\times \mathbbm{R}^{\nu}} \equiv 0$.

\medskip

\tmtextbf{-2-} Let us prove the existence statement in assertion
\ref{unicityolga6.12}. Let $p^{\tmop{conj}}$ be as in [Ha6, Theorem 2.1]. Let
$\varphi \in \mathcal{S}(\mathbbm{R}^{\nu})$. Let
\[ \psi (t, x) \assign \int_{\mathbbm{R}^{\nu}} (p^0 \times p^{\tmop{conj}})
   (t, x, y) \varphi (y) d y. \]
By (\ref{unicityolga2.5})
\[ (p^0 \times p^{\tmop{conj}}) (i \tmmathbf{t}, x, y) = (4 \pi i \Delta
   \tmmathbf{t})^{- \nu / 2} e^{- A^{- 1} (0) \cdot (x - y)^2 / 4 i
   \tmmathbf{t}} \times e^{i P_{\tmmathbf{t}}} \times k (i \tmmathbf{t})
   p^{\tmop{conj}} (i \tmmathbf{t}, x, y) \]
where the polynomial $P_{\tmmathbf{t}}$ satisfies the assumptions of Lemma
\ref{unicityolga8}. Then, by Lemma \ref{unicityolga8}, there exists $T_2 \in]
0, T]$ such that $\psi \in \mathcal{C}^{\infty} \lp{1} i] - T_2, T_2 [,
\mathcal{S}(\mathbbm{R}^{\nu}) \rp{1}$. Moreover the mapping $\varphi
\longmapsto \psi$ is continuous from $\mathcal{S}(\mathbbm{R}^{\nu})$ onto
$\mathcal{C}^{\infty} \lp{1} i] - T, T [, \mathcal{S}(\mathbbm{R}^{\nu})
\rp{1}$. Since $p^0 \times p^{\tmop{conj}}$ satisfies (\ref{unicityolga4}) and
$p^{\tmop{conj}} |_{t = 0} = 1$, (\ref{unicityolga6.14}) holds.

\medskip

\tmtextbf{-3-} Let us prove assertion \ref{unicityolga6.2}. By the regularity
properties of $p^{\tmop{conj}}$ [Ha6, Theorem 2.1], one gets a suitable
extension of the kernel of the operator $\mathcal{P}_t$. We claim that this
extension is unique. Let $p_1$ and $p_2$ be two extensions. Then
\[ p = p_1 - p_2 \in \mathcal{A} \lp{1} ( \demidisque{T} -\{0\}) \times
   \mathbbm{C}^{2 \nu} \rp{1} \cap \mathcal{C}^{\infty} \lp{1} \bar{D}^+_T
   -\{0\}, \mathcal{A}(\mathbbm{C}^{2 \nu}) \rp{1} \]
and $p|_{(i] - T, T [-\{0\}) \times \mathbbm{R}^{\nu}} = 0$. Let $(x, y) \in
\mathbbm{R}^{2 \nu}$ and let $\breve{p}$ be the function on $D_T -\{0\}$
defined by $\breve{p} (t) \assign 1_{\re t \geqslant 0} p (t, x, y)$. By
regularity properties of $p$ and Cauchy-Riemann equations with respect to $t$,
$\breve{p}$ is smooth near $i T / 2$ and satisfies Cauchy-Riemann equations.
Therefore the function $\breve{p}$ is analytic near $\text{$i T / 2$}$,
vanishes near $\text{$i T / 2$}$ and actually on $\demidisque{T} -\{0\}$. Then
by analytic continuation with respect to the space variables the function $p$
vanishes on $( \demidisque{T} -\{0\}) \times \mathbbm{C}^{2 \nu}$.

\medskip

\tmtextbf{-4-} Assertion \ref{unicityolga6.4} can also be checked by
considering the deformation formula in the free case.

\section{\label{arbitrarydirection}The heat kernel viewed from an arbitrary
direction}

We must take into account the ramification of the heat kernel at $t = 0$ in
our statements. The ramification is only due to the term $t^{- \nu / 2}$ in
(\ref{unicityolga2.5}). Let $\epsilon \in \mathbbm{R}/ 4 \pi \mathbbm{Z}$.
Then the free heat kernel $p^{\tmop{free}}$ is invariant, up to a
multiplicative constant, under the change of variables
\begin{equation}
  \label{rotatedanna2.5} (t, x, y) \mapsto (e^{i \epsilon} t, e^{i \epsilon /
  2} x, e^{i \epsilon / 2} y)
\end{equation}
and the free heat equation is invariant under the change of variables $(t, x)
\mapsto (e^{i \epsilon} t, e^{i \epsilon / 2} x)$. This elementary remark
allows a reformulation of Proposition \ref{unicityolga6} (we only consider the
harmonic case for the sake of simplicity). We denote{\footnote{Only the first
factor of the product defining $p_{\epsilon}^{\tmop{harm}}$ is concerned by
the ramification.}}
\[ p_{\epsilon}^{\tmop{harm}} \assign \left( 4 \pi e^{- i \epsilon} t
   \right)^{- \nu / 2} \times \]
\[ \left( \frac{\tmop{sh} (\omega e^{- i \epsilon} t)}{\omega e^{- i \epsilon}
   t} \right)^{- \nu / 2} \exp \lp{2} - \frac{1}{4} \frac{e^{- i
   \tmmathbf{\pi} (\epsilon)} \omega}{\tmop{sh} (\omega e^{- i \epsilon} t)}
   (\tmop{ch} (\omega e^{- i \epsilon} t) (x^2 + y^2) - 2 x \cdot y) \rp{2}
   \text{ , \ } |t| \ll 1 \text{ ,} \]
\[ p_{\epsilon}^{\tmop{free}} \assign \left( 4 \pi e^{- i \epsilon} t
   \right)^{- \nu / 2} \exp \lp{2} - \frac{1}{4 t} (x - y)^2 \rp{2} . \]
\begin{corollary}
  \label{rotatedanna4}Let $\lambda \in \mathbbm{R}$ and $\epsilon \in
  \mathbbm{R}/ 4 \pi \mathbbm{Z}$. There exists $T > 0$ such that the
  following statement holds. Let $\mu$ be a complex measure on
  $\mathbbm{R}^{\nu}$such that for every $R > 0$
  \begin{equation}
    \label{rotatedanna5} \int_{\mathbbm{R}^{\nu}} \exp (R| \xi |) d| \mu |
    (\xi) < \infty .
  \end{equation}
  Let
  \[ c (x) = \int_{\mathbbm{R}^{\nu}} \exp (ie^{- i \epsilon / 2} x \cdot \xi)
     d \mu (\xi) . \]
  Then the following assertions hold.
  \begin{enumerate}
    \item For every $\varphi \in \mathcal{S}(e^{i \epsilon / 2}
    \mathbbm{R}^{\nu})$ there exists a unique $\psi \in \mathcal{C}^{\infty}
    \lp{1} i e^{i \epsilon}] - T, T [, \mathcal{S}(e^{i \epsilon / 2}
    \mathbbm{R}^{\nu}) \rp{1}$ such that
    \[ \left\{ \begin{array}{l}
         \partial_t \psi = \lp{1} \partial_x^2 + \lambda e^{- 2 i
         \tmmathbf{\pi} (\epsilon)} x^2 + c (x) \rp{1} \psi\\
         \\
         \psi |_{t = 0} = \varphi
       \end{array} \right. \]
    and the mapping $\varphi \longmapsto \psi$ is continuous from
    $\mathcal{S}(e^{i \epsilon / 2} \mathbbm{R}^{\nu})$ onto
    $\mathcal{C}^{\infty} \lp{1} i e^{i \epsilon}] - T, T [, \mathcal{S}(e^{i
    \epsilon / 2} \mathbbm{R}^{\nu}) \rp{1}$.
    
    \item Let $p = p (t, x, y)$ be the kernel of the operator $\mathcal{P}_t :
    \mathcal{S}(e^{i \epsilon / 2} \mathbbm{R}^{\nu}) \rightarrow
    \mathcal{S}(e^{i \epsilon / 2} \mathbbm{R}^{\nu})$ defined by
    $\mathcal{P}_t (\varphi) = \psi (t, \cdot)$ for $t \in i e^{i \epsilon}] -
    T, T [$. Then $p$ can be uniquely continued as a function belonging to
    \[ \mathcal{A} \lp{1} e^{i \epsilon} ( \demidisque{T} -\{0\}) \times
       \mathbbm{C}^{2 \nu} \rp{1} \cap \mathcal{C}^{\infty} \lp{1} e^{i
       \epsilon} ( \bar{D}^+_T -\{0\}), \mathcal{A}(\mathbbm{C}^{2 \nu})
       \rp{1} . \]
    Moreover
    \[ p = p_{\epsilon}^{\tmop{harm}} \times p^{\tmop{conj}} \]
    where $p^{\tmop{conj}} = p^{\tmop{conj}} (t, x, y) \in$
    \[ \mathcal{A}(e^{i \tmmathbf{\pi} (\epsilon)} \demidisque{T} \times
       \mathbbm{C}^{2 \nu}) \cap \mathcal{C}^{\infty} \lp{1} e^{i
       \tmmathbf{\pi} (\epsilon)} \bar{D}^+_T, \mathcal{A}(\mathbbm{C}^{2
       \nu}) \rp{1} \cap \mathcal{C}^{\infty}_{b, 1} (i e^{i \tmmathbf{\pi}
       (\epsilon)}] - T, T [\times \mathbbm{R}^{2 \nu}) . \]
    \item Let us assume that $\lambda = 0$. Then
    \[ p^{\tmop{conj}} \in \mathcal{A}(e^{i \tmmathbf{\pi} (\epsilon)}
       \demiplan{} \times \mathbbm{C}^{2 \nu}) \cap \mathcal{C}^{\infty}
       \lp{1} e^{i \tmmathbf{\pi} (\epsilon)} \overline{\demiplan{}},
       \mathcal{A}(\mathbbm{C}^{2 \nu}) \rp{1} \cap \mathcal{C}^{\infty}_b (i
       e^{i \tmmathbf{\pi} (\epsilon)} \mathbbm{R} \times \mathbbm{R}^{2 \nu})
       . \]
  \end{enumerate}
\end{corollary}

\begin{figure}[h]
 \caption{}
 \centering
  \begin{pspicture}(-3,-3)(3,3.5)
\psline[linewidth=0.1mm]{->}(-1,0)(3.5,0)
\psline[linewidth=0.1mm]{->}(0,-2.7)(0,3.2)
\rput{30}(0,0){\parametricplot{0}{180}{2.5 t sin mul 2.5 t cos mul}}
\rput{30}(0,0){\psframe[linewidth=0mm,fillstyle=hlines,linecolor=white](-2,-2.9)(0,2.9)}
\rput{30}(0,0){\psline[linewidth=0.3mm](0,-2.5)(0,2.5)}
\rput{30}(0,0){\psline[linewidth=0.1mm]{->}(0,0)(4,0)}
\rput{-135}(0,0){\psline[linewidth=0.1mm]{<->}(0,0)(0,2.5)}
\uput[0](3.4,2){$e^{i\boldsymbol{\pi}(\epsilon)}\mathbbm{R}^+$}
\psarc[linewidth=0.1mm]{->}(0,0){1}{0}{30}
\uput[0](0.1,1.7){$e^{i\boldsymbol{\pi}(\epsilon)}\overline{D}_T^+$}
\uput[0](3.5,0){$\re \, t$}
\uput[0](-0.1,3.2){$\im \, t$}
\uput[0](0.8,-0.8){${T}$}
\uput[0](0.9,0.3){$\boldsymbol{\pi}(\epsilon)$}
\end{pspicture}
 \end{figure}
 
 \FloatBarrier

\begin{remark}
  Let $\epsilon \in \mathbbm{R}/ 4 \pi \mathbbm{Z}$. Let $f = f (t, x, y)$ be
  a continuous function on $\lp{1} e^{i \epsilon} ( \bar{D}^+_T -\{0\}) \rp{1}
  \times \mathbbm{C}^{\nu} \times \mathbbm{C}^{\nu}$. \ We say that $f$ goes
  to $\delta_{x = y}$ in the direction $e^{i \epsilon}$ and write
  \[ f|_{t = e^{i \epsilon} 0^+} = \delta_{x = y} \]
  if and only if for every $\theta, \vartheta \in \mathbbm{R}/ 4 \pi
  \mathbbm{Z}$, $| \theta - \epsilon | \leqslant \pi / 2$, $| \theta - 2
  \vartheta | \leqslant \pi / 2$ and every $\varphi \in \mathcal{D}(e^{i
  \vartheta} \mathbbm{R}^{\nu})$, $x \in e^{i \vartheta} \mathbbm{R}^{\nu}$
  \[ \int_{e^{i \vartheta} \mathbbm{R}^{\nu}} f (r e^{i \theta}, x, y) \varphi
     (y) d y \longrightarrow_{r \rightarrow 0^+} \varphi (x) . \]
  Here $d y = e^{i \vartheta \nu} d m (y)$ where $m$ denotes the standard (non
  negative) Lebesgue measure on $e^{i \vartheta} \mathbbm{R}^{\nu}$. Then the
  kernel $p$ satisfies on $e^{i \epsilon} ( \demidisque{T} -\{0\}) \times
  \mathbbm{C}^{2 \nu}$
  \begin{equation}
    \label{rotatedanna6} \left\{ \begin{array}{l}
      \partial_t p = \lp{1} \partial_x^2 + \lambda e^{- 2 i \tmmathbf{\pi}
      (\epsilon)} x^2 + c (x) \rp{1} p\\
      \\
      p|_{t = e^{i \epsilon} 0^+} = \delta_{x = y}
    \end{array} \right.
  \end{equation}
  since, for every smooth function $g = g (t, x, y)$ on $e^{i \tmmathbf{\pi}
  (\epsilon)} \bar{D}^+_T \times \mathbbm{C}^{\nu} \times \mathbbm{C}^{\nu}$
  such that $g|_{t = 0} = 1$, the function $p_{\epsilon}^{\tmop{free}} \times
  g$ goes to $\delta_{x = y}$ in the direction $e^{i \epsilon}$.
\end{remark}

Let us choose $\epsilon = \pi$. Then we get a solution $p$ such that
$p^{\tmop{conj}}$ is defined on
\[ \lbc{1} t = |t|e^{i \theta} \in \mathbbm{C}| \theta \in [\pi / 2, 3 \pi /
   2]_{4 \pi}, |t| < T \rbc{1} \times \mathbbm{C}^{2 \nu} \]
or $\lbc{1} t \in \mathbbm{C}| \theta \in [\pi / 2, 3 \pi / 2]_{4 \pi} \rbc{1}
\times \mathbbm{C}^{2 \nu}$ if $\lambda = 0$. In particular, by considering
values of $t$ such that $\arg t = \pi / 2, 3 \pi / 2$, we obtain the following
result about the standard Schr\"odinger equation.

\begin{corollary}
  \label{rotatedanna6.1}Let $\mu$ as in Corollary \ref{rotatedanna4}. Let
  \begin{equation}
    \label{rotatedanna6.5} c (x) = \int_{\mathbbm{R}^{\nu}} \exp (x \cdot \xi)
    d \mu (\xi) .
  \end{equation}
  Let $\lambda \in \mathbbm{R}$. Then there exist $T > 0$ and
  \[ p^{\tmop{conj}} = p^{\tmop{conj}} ( \tmmathbf{t}, x, y) \in
     \mathcal{C}^{\infty} \lp{1}] - T, T [, \mathcal{A}(\mathbbm{C}^{2 \nu})
     \rp{1} \]
  such that $p = p_{\epsilon}^{\tmop{harm}} \times p^{\tmop{conj}}$ satisfies
  \begin{equation}
    \label{rotatedanna8} \left\{ \begin{array}{l}
      \frac{1}{i} \partial_{\tmmathbf{t}} p = \lp{1} \partial_x^2 + \lambda
      x^2 + c (x) \rp{1} p \text{ , \ } x \in \mathbbm{R}^{\nu}, \tmmathbf{t}
      \in] - T, T [\\
      \\
      p|_{\tmmathbf{t} = 0} = \delta_{x = y} \text{ , \ } y \in
      \mathbbm{R}^{\nu}
    \end{array} . \right.
  \end{equation}
  If $\lambda = 0$, $p^{\tmop{conj}} \in \mathcal{C}^{\infty} \lp{1}
  \mathbbm{R}, \mathcal{A}(\mathbbm{C}^{2 \nu}) \rp{1}$.
\end{corollary}

\begin{remark}
  \label{rotatedanna8.5} The assumption (\ref{rotatedanna5}), since $c$ is
  given by (\ref{rotatedanna6.5}), allows potentials such as
  \[ V (x) = \lambda x^2 \pm e^{x^2}, V (x) = \lambda x^2 \pm e^{x_1}, \ldots
  \]
  In the case $\lambda = 0$, this fact was noticed by Kuna, Streit and
  Westerkamp [K-S-W]. In particular the function $x \longmapsto e^{x^2}$ is
  viewed as a perturbation of the operator $\partial_x^2 + \lambda x^2$ (the
  deformation formula is used to deal with this part of the potential $V$)
  whereas the function $x \longmapsto \lambda x^2$ is not viewed as a
  perturbation of the operator $\partial_x^2$ in our method. Using a complex
  point of view with respect to the space variables perhaps explains this
  ``paradox''. A related remark can be done for the uniqueness problem: we do
  not claim that (\ref{rotatedanna8}) has a unique solution in its natural
  real setting. However it does, if we consider complex values for the space
  variables ($x, y \in e^{i \pi / 2} \mathbbm{R}^{\nu}$), by taking advantage
  of the uniqueness statement of Corollary \ref{rotatedanna4}.
  
  Notice that the dilation given by $(t, x, y) \mapsto (e^{i \pi} t, i x, i
  y)$ ($\epsilon = \pi$ in \ref{rotatedanna2.5}), which allows one to view
  Corollary \ref{rotatedanna6.1} as a consequence of Corollary
  \ref{rotatedanna4}, ``reverses'' the direction of $t$.
\end{remark}

Another viewpoint is formally related to the previous proposition. For $\theta
> 0$, let
\[ \mathbbm{R}^{+, \nu}_{\prec, \theta} \assign \{e^{i \varphi} x \in \mathbbm{C}^{\nu}|x \in \lp{0}
   \mathbbm{R}^+ \srp{0}{\nu}{}, \varphi \in] - \theta, \theta [_{2 \pi} \},
\]
\[ \mathbbm{R}^{\nu}_{\prec, \theta} \assign \{e^{i \varphi} x \in \mathbbm{C}^{\nu}|x \in
   \mathbbm{R}^{\nu}, \varphi \in] - \theta, \theta [_{2 \pi} \}, \]
\[ \mathbbm{R}^+_{\prec, \theta} \assign \{r e^{i \varphi} \in \mathbbm{C}|r >
   0, \varphi \in] - \theta, \theta [_{2 \pi} \}. \]
One has

\begin{proposition}
  \label{rotatedanna9}Let $\theta, \alpha \in] 0, \pi / 4 [$. Let $\mu$ be a
  $\mathbbm{C}$-valued Borel measure on $\mathbbm{C}^{\nu}$. Let us assume
  (case $1$) that $\mathbbm{R}^{+, \nu}_{\prec, \theta}$ contains the support
  of $\mu$ and that
  \[ \forall R > 0, \int_{\mathbbm{\mathbbm{C}}^{\nu}} \exp (R| \xi |) d| \mu
     | (\xi) < \infty \]
  or (case $2$) that $d \mu (\xi) = \hat{c} (\xi) d \xi$ where $\hat{c}$
  denotes an analytic function on $\mathbbm{R}^{\nu}_{\prec, \alpha}$
  satisfying
  \[ \forall R > 0, \exists K > 0, \forall \xi \in
     \text{$\mathbbm{R}^{\nu}_{\prec, \alpha}$}, | \hat{c} (\xi) | \leqslant K
     e^{- R| \xi |} . \]
  Let $F =\mathbbm{C}^{\nu}$ (case $1$) or $F =\mathbbm{R}^{\nu}$ (case $2$).
  Let
  \[ c (x) \assign \int_F \exp (ix \cdot \xi) d \mu (\xi) . \]
  Then, by the deformation formula,
  \begin{itemizeminus}
    \item case $1$: The heat equation associated to the operator $\partial_x^2
    + c (x)$ has a solution $p$ defined on $\lp{1} \mathbbm{R}^+_{\prec, \pi /
    2 - 2 \theta} -\{0\} \rp{1} \times \mathbbm{C}^{2 \nu}$ satisfying the
    following boundary condition. For every $\varphi \in
    \mathcal{D}(\mathbbm{R}^{\nu})$, $x \in \mathbbm{R}^{\nu}$ and $\alpha
    \in] - \pi / 2 + 2 \theta, \pi / 2 - 2 \theta [$
    \[ \int_{\mathbbm{R}^{\nu}} p (r e^{i \alpha}, x, y) \varphi (y) d y
       \longrightarrow_{r \rightarrow 0^+} \varphi (x) . \]
    \item case $2$: The heat kernel of the operator $\partial_x^2 + c (x)$,
    which is defined on $\mathbbm{R}^+ \times \mathbbm{R}^{2 \nu}$, admits an
    analytic continuation on $\lp{1} \mathbbm{R}^+_{\prec, \pi / 2 + 2 \alpha}
    -\{0\} \rp{1} \times \mathbbm{C}^{2 \nu}$.
  \end{itemizeminus}
\end{proposition}

\begin{figure}[h]
\caption{}
\begin{pspicture}(-2,-3)(2,3.5)
\psset{unit=0.9}
\pswedge[fillstyle=hlines,linecolor=white]{2}{60}{300}
\psline[linewidth=0.1mm]{->}(-1,0)(2.7,0)
\psline[linewidth=0.1mm]{->}(0,-1.7)(0,2.5)
\rput{60}(0,0){\psline[linewidth=0.3mm](0,0)(3,0)}
\rput{-60}(0,0){\psline[linewidth=0.3mm](0,0)(3,0)}
\uput[0](0.8,-1.1){$\mathbbm{R}^+_{\prec,\frac{\pi}{2}-2\theta}$}
\psarc[linewidth=0.1mm]{->}(0,0){0.5}{0}{60}
\uput[0](2.7,0){$\re \, t$}
\uput[0](-0.1,2.7){$\im \, t$}
\uput[0](0.4,0.3){$\frac{\pi}{2}-2\theta$}
\uput[0](-0.45,-2.3){case $1$}
\rput{0}(8,0){\pswedge[fillstyle=hlines,linecolor=white]{2}{150}{210}}
\rput{0}(8,0){\psline[linewidth=0.1mm]{->}(-1,0)(2,0)}
\rput{0}(8,0){\psline[linewidth=0.1mm]{->}(0,-0.7)(0,2.5)}
\rput{0}(8,0){\rput{150}(0,0){\psline[linewidth=0.3mm](0,0)(3,0)}}
\rput{0}(8,0){\rput{210}(0,0){\psline[linewidth=0.3mm](0,0)(3,0)}}
\rput{0}(8,0){\uput[0](-0.5,-1.1){$\mathbbm{R}^+_{\prec,\frac{\pi}{2}+2\alpha}$}}
\rput{0}(8,0){\psarc[linewidth=0.1mm]{->}(0,0){0.5}{0}{150}}
\rput{0}(8,0){\uput[0](2,0){$\re \, t$}}
\rput{0}(8,0){\uput[0](-0.1,2.7){$\im \, t$}}
\rput{0}(8,0){\uput[0](0,0.7){$\frac{\pi}{2}+2\alpha$}}
\rput{0}(8,0){\uput[0](-0.45,-2.3){case $2$}}
\end{pspicture}
\end{figure}
\FloatBarrier
\begin{proof}
  Let $p^{\tmop{conj}} \assign \sum_{n \geqslant 0} v_n$ where
  \[ v_n \assign t^n \int_{0 < s_1 < \cdots < s_n < 1} \int_F e^{i \lp{1} y +
     s (x - y) \rp{1} \cdot \xi} \exp \lp{1} - t s (1 - s) \cdot_n \xi \otimes
     \xi \rp{1} d^{\nu n} \mu^{\otimes} (\xi) d^n s, \]
  \[ \lp{1} y + s (x - y) \rp{1} \cdot \xi : = \lp{1} y + s_1 (x - y) \rp{1}
     \cdot \xi_1 + \cdots + \lp{1} y + s_n (x - y) \rp{1} \cdot \xi_n, \]
  \[ s (1 - s) \cdot_n \xi \otimes \xi : = \sum^n_{j, k = 1} s_{j \wedge k} (1
     - s_{j \vee k}) \xi_j \cdot \xi_k, \]
  \[ d^{\nu n} \mu^{\otimes} (\xi) \assign d \mu (\xi_n) \cdots d \mu (\xi_1)
     . \]
  We first check that the series defining $p^{\tmop{conj}}$ is convergent.
  
  (case 1) We claim that $p^{\tmop{conj}} = p^{\tmop{conj}} (t, x, y) \in
  \mathcal{A}(\mathbbm{R}^+_{\prec, \pi / 2 - 2 \theta} \times \mathbbm{C}^{2
  \nu})$. One has
  \[ |x|, |y| \leqslant R \Rightarrow \ve{1} \exp \lp{1} i \lp{1} y + s (x -
     y) \rp{1} \cdot \xi \rp{1} \ve{1} \leqslant e^{R| \xi_1 |} \times \cdots
     \times e^{R| \xi_n |} . \]
  Since $2 \theta < \pi / 2$, $\mathbbm{R}^+_{\prec, 2 \theta}$ is a convex
  cone. Then
  \[ t \in \mathbbm{R}^+_{\prec, \pi / 2 - 2 \theta}, \xi \in \tmop{supp}
     (\mu^{\otimes}) \Rightarrow \re \lp{1} t s (1 - s) \cdot_n \xi \otimes
     \xi \rp{1} \geqslant 0. \]
  This implies the convergence of the series defining $p^{\tmop{conj}}$ and
  the analyticity of $p^{\tmop{conj}}$.
  
  (case $2$) We claim that $p^{\tmop{conj}} \in
  \mathcal{A}(\mathbbm{R}^+_{\prec, \pi / 2 + 2 \alpha} \times \mathbbm{C}^{2
  \nu})$. Let $\beta \in] 0, \alpha [$. Since the function $\hat{c}$ is
  analytic on $\mathbbm{R}^{\nu}_{\prec, \alpha}$, one gets by a deformation
  of the integration contour,
  \[ v_n \assign e^{- i \nu n \beta} t^n \int_{0 < s_1 < \cdots < s_n < 1}
     \int_{\mathbbm{R}^{\nu n}} e^{i e^{- i \beta} \lp{1} y + s (x - y) \rp{1}
     \cdot \xi} \]
  \[ \exp \lp{1} - t e^{- 2 i \beta} s (1 - s) \cdot_n \xi \otimes \xi \rp{1}
     \hat{c} (e^{- i \beta} \xi_1) \cdots \hat{c} (e^{- i \beta} \xi_n) d^{\nu
     n} \xi d^n s. \]
  Therefore the convergence of the series defining $p^{\tmop{conj}}$ and the
  analyticity of $p^{\tmop{conj}}$ hold for $\re \lp{1} e^{- 2 i \beta} t
  \rp{1} > 0$ and $x, y \in \mathbbm{C}^{\nu}$. Since $\beta$ is arbitrary,
  one gets that $p^{\tmop{conj}} \in \mathcal{A}(\mathbbm{R}^+_{\prec, \pi / 2
  + 2 \alpha} \times \mathbbm{C}^{2 \nu})$.
  
  By proceeding as in [Ha4], one can show that $p = p^{\tmop{free}} \times
  p^{\tmop{conj}}$ satisfies the heat equation. Moreover the boundary
  condition is satisfied.
\end{proof}

\begin{example}
  Let $\theta_1, \ldots, \theta_q \in] - \pi / 4, \pi / 4 [$ and $\lambda_1,
  \ldots, \lambda_q \in (\mathbbm{R}^+)^{\nu}$. Let
  \[ c (x) = \exp \lp{1} i e^{i \theta_1} \lambda_1 \cdot x \rp{1} + \cdots +
     \exp \lp{1} i e^{i \theta_q} \lambda_q \cdot x \rp{1} . \]
  Then the function $c$ satisfies the assumptions of Proposition
  \ref{rotatedanna9} (case $1$). We do not attempt to give a uniqueness
  statement in this case.
\end{example}

\begin{example}
  Let $c (x) = e^{- x^2}$. By Proposition \ref{rotatedanna9} (case $2$) the
  heat kernel is well defined on $(\mathbbm{C}-] - \infty, 0]) \times
  \mathbbm{C}^{2 \nu}$.
\end{example}

\begin{remark}
  One can generalize Proposition \ref{rotatedanna9} (case $2$) in the harmonic
  case. The proof needs a modification of [Ha4, Lemma 4.2].
\end{remark}

\section{\label{borelsummation}Borel summability of the conjugate of the heat
kernel in an arbitrary direction}

For the sake of simplicity, we only consider the free case in this section.
Let $\kappa, T > 0$. Let
\[ \cylindre{\kappa} \assign \lbc{1} z \in \mathbbm{C} | d (z, [0, + \infty [)
   < \kappa \rbc{1} \text{ , \ } \nev{T} \assign \lbc{2} z \in \mathbbm{C} |
   \re \lp{1} \frac{1}{z} \rp{1} > \frac{1}{T} \rbc{2} . \]
$\nev{T}$ is the open disk of center $\frac{T}{2}$ and radius $\frac{T}{2}$.

\begin{definition}
  Let $\dot{\epsilon} \in \mathbbm{R}/ 2 \pi \mathbbm{Z}$. Let $a_1, \ldots,
  a_r, \ldots \in \mathbbm{C}$. The formal power series $\tilde{f} = \sum_{r
  \geqslant 0} a_r t^r$ is called Borel-Nevalinna (respectively Borel-Watson)
  summable in the direction $e^{i \dot{\epsilon}}$ if
  \begin{itemize}
    \item the radius of convergence of the Borel transform of $\tilde{f}$,
    $\hat{f} (\tau) \assign \sum_{r = 0}^{\infty} \frac{a_r}{r!} \tau^r$, does
    not vanish
    
    \item there exist $\kappa > 0$ (respectively $\theta > 0$) such that the
    Borel transform can be analytically continued on $e^{i \dot{\epsilon}}
    \cylindre{\kappa}$ (respectively $e^{i \dot{\epsilon}}
    \mathbbm{R}^+_{\prec, \theta}$)
    
    \item there exist $K, T > 0$ such that for every $\tau \in \text{$e^{i
    \dot{\epsilon}} \cylindre{\kappa}$}$ (respectively $e^{i \dot{\epsilon}}
    \mathbbm{R}^+_{\prec, \theta}$)
    \[ | \hat{f} (\tau) | \leqslant Ke^{^{} | \tau | / T} . \]
    If the power series $\tilde{f}$ is Borel-Nevalinna or Borel-Watson
    summable, the Laplace transform of $\hat{f}$
    \[ f (t) \assign \int_0^{+ \infty} \hat{f} (\tau) e^{- \frac{\tau}{t}}
       \frac{d \tau}{t} \]
    is called the Borel sum of $\tilde{f}$.
  \end{itemize}
\end{definition}

\begin{figure}[h]
\caption{}
\begin{pspicture}(-3,-1)(3,5)
\psline[linewidth=0.1mm]{->}(-1,0)(5.5,0)
\psline[linewidth=0.1mm]{->}(0,-1.2)(0,3.2)
\rput{30}(0,0){\parametricplot[linewidth=0.1mm]{180}{360}{0.7 t sin mul 0.7 t cos mul}}
\rput{30}(0,0){\psline[linewidth=0.1mm]{->}(0,0)(6,0)}
\rput{30}(0,0){\psline[linewidth=0.1mm](0,0.7)(5,0.7)}
\rput{30}(0,0){\psline[linewidth=0.1mm](0,-0.7)(5,-0.7)}
\rput{50}(0,0){\psline[linewidth=0.1mm](0,0)(6,0)}
\rput{10}(0,0){\psline[linewidth=0.1mm](0,0)(6,0)}
\rput{-220}(0,0){\psline[linewidth=0.1mm]{<->}(0,0)(0.7,0)}
\uput[0](5.2,3.1){$e^{i\dot \epsilon}\mathbbm{R}^+$}
\uput[0](2.3,2.1){$e^{i\dot \epsilon}{\tilde{S}}_\kappa$}
\uput[0](3.3,3.9){$e^{i\dot \epsilon}\mathbbm{R}^+_{\prec,\theta}$}
\psarc[linewidth=0.1mm]{->}(0,0){1}{0}{30}
\psarc[linewidth=0.1mm]{->}(0,0){1.5}{30}{50}
\uput[0](5.5,0){$\re \, \tau$}
\uput[0](-0.1,3.2){$\im \, \tau$}
\uput[0](-0.45,0.35){${\kappa}$}
\uput[0](0.9,0.35){$\dot \epsilon$}
\uput[0](1.12,1.12){$\theta$}
\end{pspicture}
\end{figure}
\FloatBarrier

\begin{figure}[h]
\caption{}
\centering
\begin{pspicture}(-3,-2.8)(3,4)
\psset{unit=0.9}
\psline[linewidth=0.1mm]{->}(-1.5,0)(4.5,0)
\psline[linewidth=0.1mm]{->}(0,-3.7)(0,3.7)
\pswedge[fillstyle=hlines,linecolor=white]{3.5}{130}{290}
\rput{30}(0,0){\parametricplot{-10}{190}{3 t sin mul 3 t cos mul}}
\rput{30}(0,0){\parametricplot{0}{380}{1.5 1.5 t sin mul add 1.5 t cos mul}}
\rput{-35}(0,0){\psline[linewidth=0.1mm]{<->}(0,0)(3,0)}
\rput{30}(0,0){\psline[linewidth=0.1mm]{->}(0,0)(5.5,0)}
\rput{-60}(0,0){\psline[linewidth=0.1mm,linestyle=dashed](-3.5,0)(3.5,0)}
\rput{130}(0,0){\psline[linewidth=0.3mm](0,0)(3,0)}
\rput{-70}(0,0){\psline[linewidth=0.3mm](0,0)(3,0)}
\psarc[linewidth=0.1mm]{->}(0,0){1}{0}{30}
\psarc[linewidth=0.1mm]{->}(0,0){0.5}{30}{130}
\uput[0](0.8,1.5){$e^{i\dot \epsilon}\nev{T}$}
\uput[0](4.5,0){$\re \, t$}
\uput[0](-0.1,4){$\im \, t$}
\uput[0](4.8,2.8){$e^{i\dot \epsilon}\mathbbm{R}^+$}
\uput[0](0.9,0.35){$\dot \epsilon$}
\uput[0](-0.1,0.65){$\theta$}
\uput[0](-1.33,2.35){$e^{i\dot \epsilon}\mathbbm{R}^+_{\prec,\theta}\cap D_T$}
\end{pspicture}
\end{figure}
\FloatBarrier

\begin{remark}
  If a power series $\tilde{f}$ is Borel-Nevalinna (respectively Borel-Watson)
  summable in the direction $e^{i \dot{\epsilon}}$, then there exist $T > 0$
  and $\theta > \pi / 2$ such that its Borel sum is well defined for $t \in
  e^{i \dot{\epsilon}} \nev{T}$ (respectively $e^{i \dot{\epsilon}}
  \mathbbm{R}^+_{\prec, \theta} \cap D_T$). See [So].
\end{remark}
The change of variables (\ref{rotatedanna2.5}) allows us to give the following
corollary of Theorem 3.1 [Ha4].

\begin{corollary}
  \label{borelemilia2}Let $\varepsilon > 0$ and $\epsilon \in \mathbbm{R}/ 4
  \pi \mathbbm{Z}$. Let $\mu$ be a $\mathbbm{C}$-valued measure on
  $\mathbbm{R}^{\nu}$ verifying
  \begin{equation}
    \label{borelemilia2.5} \int_{\mathbbm{R}^{\nu}} \exp (\varepsilon \xi^2)
    d| \mu | (\xi) < \infty .
  \end{equation}
  Let
  \begin{equation}
    \label{borelemilia3} c (x) = \int \exp (ie^{- i \epsilon / 2} x \cdot \xi)
    d \mu (\xi)
  \end{equation}
  and let $u$ be the solution of (\ref{rotatedanna6}) where $\lambda = 0$. Let
  $p^{\tmop{conj}}$ be defined by u=$p^{\tmop{free}} p^{\tmop{conj}}$. Then
  $p^{\tmop{conj}}$ admits a Borel transform $\widehat{p^{\tmop{conj}}}$ (with
  respect to $t$) which is analytic on $\mathbbm{C}^{1 + 2 \nu}$. Let $\kappa,
  R > 0$ and let
  \[ C \assign 2 \lp{2} \int \exp \lp{1} \frac{2 \kappa}{\varepsilon} +
     \frac{\varepsilon}{2} \xi^2 + R| \xi | \rp{1} d| \mu | (\xi) \srp{2}{1 /
     2}{} . \]
  Then, for every $(\tau, x, y) \in e^{i \tmmathbf{\pi} (\epsilon)}
  \tilde{S}_{\kappa} \times \mathbbm{C}^{2 \nu}$ such that $\ve{1} \mathcal{I}
  m (e^{- i \epsilon / 2} x) \ve{1} < R$ and $\ve{1} \mathcal{I} m (e^{- i
  \epsilon / 2} y) \ve{1} < R$,
  \begin{equation}
    \label{borelemilia4} \ve{1} \widehat{p^{\tmop{conj}}} (\tau, x, y) \ve{1}
    \leqslant \exp \lp{1} C| \tau |^{1 / 2} \rp{1} .
  \end{equation}
\end{corollary}

\begin{remark}
  By the estimate (\ref{borelemilia4}), the small time expansion of the
  conjugate heat kernel is Borel-Nevalinna summable in the direction $e^{i
  \tmmathbf{\pi} (\epsilon)}$ and its Borel sum is equal to $p^{\tmop{conj}}$.
\end{remark}

We now illustrate Corollary \ref{borelemilia2} by simple examples.

\begin{example}
  Let $\epsilon \in \mathbbm{R}/ 4 \pi \mathbbm{Z}$, $\xi_0 \in e^{- i
  \epsilon / 2} \mathbbm{R}^{\nu} -\{0\}$ and $c (x) = \exp (ix \cdot \xi_0)$.
  The function $c$ satisfies the assumptions of Corollary \ref{borelemilia2}
  hence the small time expansion of the conjugate heat kernel is
  Borel-Nevalinna summable in the direction $e^{i \tmmathbf{\pi} (\epsilon)}$.
  For $\epsilon' \in \mathbbm{R}/ 4 \pi \mathbbm{Z}$, $\tmmathbf{\pi}
  (\epsilon') \neq \tmmathbf{\pi} (\epsilon)$, the function $c$ is not bounded
  on $e^{i \epsilon' / 2} \mathbbm{R}^{\nu}$ and therefore does not satisfy
  the assumptions of Corollary \ref{borelemilia2} in the direction $e^{i
  \varepsilon'}$: Corollary \ref{borelemilia2} can not be used to study the
  Borel-Watson summability of this expansion in the direction $e^{i
  \tmmathbf{\pi} (\epsilon)}$. 
\end{example}

\begin{example}
  Let $c (x) = \exp \lp{1} i x_1 + i e^{i \pi / 8} x_2 \rp{1}$ and let $p$ be
  the solution given by Proposition \ref{rotatedanna9} (case $1$). Then
  Corollary \ref{borelemilia2} can not help us to study the Borel summability
  of the small time expansion of $p$.
  
  Let us now consider $c (x) = \exp \lp{1} i x_1 \rp{1} + \exp \lp{1} i e^{i
  \pi / 8} x_2 \rp{1}$. Then by separation of variables, the solution given by
  Proposition \ref{rotatedanna9} is the product of two Borel-Nevalinna
  summable expansions but in different directions.
\end{example}

\begin{example}
  Let $K, A > 0$ and $\alpha \in] 0, \pi / 2 [$. Let $\hat{c}$ be an
  analytical function on $\mathbbm{R}^{\nu}_{\prec, \alpha}$ satisfying, for
  every $\xi \in \mathbbm{R}^{\nu}_{\prec, \alpha}$,
  \[ | \hat{c} (\xi) | \leqslant K e^{- A| \xi |^2} . \]
  Let
  \[ c (x) \assign \int_{\mathbbm{R}^{\nu}} \exp (ix \cdot \xi) \hat{c} (\xi)
     d \xi . \]
  Let $\epsilon \in I \assign] - 2 \alpha, 2 \alpha [_{4 \pi}$ and
  $\varepsilon < A$. Then there exists a measure $\mu$ satisfying
  (\ref{borelemilia2.5}) such that the function $c$ is also defined by
  (\ref{borelemilia3}) (see also Proposition \ref{rotatedanna9} case $2$).
  Therefore the small time expansion of the conjugate heat kernel is
  Borel-Nevalinna or Borel-Watson summable in every direction belonging to
  $\tmmathbf{\pi} (I)$. Functions like $c = e^{- \gamma x^2}$, $\gamma > 0$,
  satisfy such a property.
\end{example}

\section{Appendix}

Here is a proof of Lemma \ref{unicityolga8}.

For the sake of simplicity, we assume $B = \tmop{Id}$. For $\delta \in
\mathbbm{N}^{\nu}$, we denote $| \delta | = \delta_1 + \cdots + \delta_{\nu}$.
For $m, k \in \mathbbm{R}$ and $p \in \mathbbm{N}$, we denote by
\[ S_p^{m, k} (] - T, T [\times \mathbbm{R}^{\nu} \times \mathbbm{R}^{\nu})
\]
the set of smooth functions $f = f ( \tmmathbf{t}, x, y)$ on $] - T, T [\times
\mathbbm{R}^{\nu} \times \mathbbm{R}^{\nu}$ such that
\[ \forall (q, r) \in \mathbbm{N}^2, \exists C > 0, \forall (\alpha, \beta,
   \gamma) \in \mathbbm{N} \times \mathbbm{N}^{\nu} \times \mathbbm{N}^{\nu},
   \forall ( \tmmathbf{t}, x, y) \in] - T, T [\times \mathbbm{R}^{\nu} \times
   \mathbbm{R}, \]
\[ \alpha \leqslant p, | \beta | \leqslant q, | \gamma | \leqslant r
   \Rightarrow | \partial_{\tmmathbf{t}}^{\alpha} \partial^{\beta}_x
   \partial^{\gamma}_y f| \leqslant C (1 + |x|)^{m + \alpha} (1 + |y|)^{k +
   \alpha} . \]
For such a function, we denote by $|f|_{m, k, p, q, r}$ the best constant $C$
satisfying the previous inequality. For $f \in S_p^{m, k}$ ($k < - \nu$), let
us denote
\[ \mathcal{F}f ( \tmmathbf{t}, x) \assign \int_{\mathbbm{R}^{\nu}} (4 \pi i
   \tmmathbf{t})^{- \nu / 2} e^{- (x - y)^2 / 4 i \tmmathbf{t}} e^{i \nonesep
   P_{\tmmathbf{t}} (x, y)} f ( \tmmathbf{t}, x, y) d y. \]
We first establish some useful properties of this transform. Let $f \in
S_r^{m, k}$. Let $j = 1, \ldots, \nu$.
\begin{itemize}
  \item Using the symmetry of the free Schr\"odinger kernel and integration by
  parts, one gets
  \begin{equation}
    \label{appendixsonia2} \partial_{x_j} \mathcal{F}f =\mathcal{F} \tilde{f}
  \end{equation}
  where
  \[ \tilde{f} \assign e^{- i \nonesep P_{\tmmathbf{t}} (x, y)}
     (\partial_{x_j} + \partial_{y_j}) \lp{1} e^{i \nonesep P_{\tmmathbf{t}}
     (x, y)} f ( \tmmathbf{t}, x, y) \rp{1} \in S_{r - 1}^{m + 1, k + 1} . \]
  Moreover there exists $c_2 > 0$, which only depends on the coefficients of
  $P_{\tmmathbf{t}} (x, y)$, such that
  \[ | \tilde{f} |_{m + 1, k + 1, p, q - 1, r - 1} \leqslant c_2 |f|_{m, k,
     p, q, r} . \]
  \item Since
  \[ \frac{1}{i} \partial_{\tmmathbf{t}} \lp{1} (4 \pi i \tmmathbf{t})^{- \nu
     / 2} e^{- (x - y)^2 / 4 i \tmmathbf{t}} \rp{1} = \partial^2_y \lp{1} (4
     \pi i \tmmathbf{t})^{- \nu / 2} e^{- (x - y)^2 / 4 i \tmmathbf{t}} \rp{1}
  \]
  and by integrations by parts, one gets
  \begin{equation}
    \label{appendixsonia4} \partial_{\tmmathbf{t}} \mathcal{F}f =\mathcal{F}
    \tilde{f}
  \end{equation}
  where
  \[ \tilde{f} \assign e^{- i \nonesep P_{\tmmathbf{t}} (x, y)} (i
     \partial^2_y + \partial_{\tmmathbf{t}}) \lp{1} e^{i \nonesep
     P_{\tmmathbf{t}} (x, y)} f ( \tmmathbf{t}, x, y) \rp{1} \in S_{r - 2}^{m
     + 2, k + 2} . \]
  Moreover there exists $c_1 > 0$ such that
  \[ | \tilde{f} |_{m + 2, k + 2, p - 1, q, r - 2} \leqslant c_1 |f|_{m, k, p,
     q, r} . \]
  \item We shall need to estimate $x_j \mathcal{F}f$. For this, we express the
  multiplication operator by $x_j$ in a convenient way. Let us denote
  \[ \phi \assign \frac{(x - y)^2}{4 \tmmathbf{t}} + P_{\tmmathbf{t}} (x, y) .
  \]
  Let $\varsigma = 1, \ldots, \nu$. Then
  \[ \partial_{y_{\varsigma}} \phi = \frac{y_{\varsigma} - x_{\varsigma}}{2
     \tmmathbf{t}} + \frac{1}{2} c ( \tmmathbf{t}) + \frac{1}{2}
     \sum_{\varsigma' = 1}^{\nu} \lp{1} a_{\varsigma, \varsigma'} (
     \tmmathbf{t}) x_{\varsigma'} + b_{\varsigma, \varsigma'} ( \tmmathbf{t})
     y_{\varsigma'} \rp{1} \]
  where $a_{\varsigma, \varsigma'}, b_{\varsigma, \varsigma'}, c$ are smooth
  $\mathbbm{R}$-valued functions defined on $] - T, T [$. Then
  \[ \lp{1} 1 - \tmmathbf{t} a_{\varsigma, \varsigma} ( \tmmathbf{t}) \rp{1}
     x_{\varsigma} - \tmmathbf{t} \sum_{\tmscript{\begin{array}{l}
       \varsigma' = 1\\
       \varsigma' \neq \varsigma
     \end{array}}}^{\nu} a_{\varsigma, \varsigma'} ( \tmmathbf{t})
     x_{\varsigma'} = y_{\varsigma} + \tmmathbf{t} c ( \tmmathbf{t}) - 2
     \tmmathbf{t} \partial_{y_{\varsigma}} \phi + \tmmathbf{t}
     \sum_{\varsigma' = 1}^{\nu} b_{\varsigma, \varsigma'} ( \tmmathbf{t})
     y_{\varsigma'} . \]
  Let us consider the above equations as a system of $\nu$ equations where the
  unknowns are $x_1, \ldots, x_{\nu}$. Then there exists $T_2 \in] 0, T [$
  such that, for $\tmmathbf{t} \in] - T_2, T_2 [$,
  \[ x_j = u ( \tmmathbf{t}) \cdot \partial_y \phi + v ( \tmmathbf{t}) \cdot y
     + w ( \tmmathbf{t}) \]
  where $u, v$ (respectively $w$) are smooth $\mathbbm{R}^{\nu}$-valued
  (respectively $\mathbbm{R}$-valued) functions defined on $] - T_2, T_2 [$.
  These functions and $T_2$ only depend on the coefficients of the polynomial
  $P_{\tmmathbf{t}}$. Then, by integration by parts,
  \begin{equation}
    \label{appendixsonia6} x_j \mathcal{F}f \assign \mathcal{F} \tilde{f}
  \end{equation}
  where
  \[ \tilde{f} \assign i u ( \tmmathbf{t}) \cdot \partial_y f + \lp{1} v (
     \tmmathbf{t}) \cdot y + w ( \tmmathbf{t}) \rp{1} f \in S_{r - 1}^{m, k +
     1} . \]
  Moreover there exists $c_3 > 0$ such that
  \[ | \tilde{f} |_{m, k + 1, p, q, r - 1} \leqslant c_3 |f|_{m, k, p, q, r}
     . \]
\end{itemize}

\medskip

\tmtextbf{-} Let $k \in \mathbbm{R}$ and $r \in \mathbbm{N}$. Let $\varphi \in
\mathcal{S}(\mathbbm{R}^{\nu})$ and $u \in \mathcal{C}^{\infty}_{b, 1} (] - T,
T [\times \mathbbm{R}^{2 \nu})$. Let us denote
\[ \| \varphi \|_{k, r} \assign \sup_{| \gamma | \leqslant r, y \in
   \mathbbm{R}^{\nu}} (1 + |y|)^{- k} | \partial^{\gamma}_y \varphi |. \]
By Leibniz formula, the function
\[ f : ( \tmmathbf{t}, x, y) \longmapsto u ( \tmmathbf{t}, x, y) \varphi (y)
\]
belongs to $S_r^{0, k}$ and for every $p, q \in \mathbbm{N}$, there exists $C
> 0$ such that
\[ |f|_{0, k, p, q, r} \leqslant C\| \varphi \|_{k, r} \]
($C$ depends on the function $u$ and the numbers $k, p, q, r$). Let
$\tmmathbf{\psi}$ be defined by (\ref{unicityolga8.1}). Then $\tmmathbf{\psi}
=\mathcal{F}f$. Let $(\alpha, \beta, \delta) \in \mathbbm{N} \times
\mathbbm{N}^{\nu} \times \mathbbm{N}^{\nu}$. Let us assume that
\begin{equation}
  \label{appendixsonia8} \left\{ \begin{array}{l}
    k + 2 \alpha + | \beta | + | \delta | < \um \nu - 1\\
    \\
    p - \alpha \geqslant 0 \text{ , \ } q - | \beta | \geqslant 0 \text{ , \ }
    r - 2 \alpha - | \beta | - | \delta | \geqslant 0
  \end{array} \text{} . \right.
\end{equation}
Then, by (\ref{appendixsonia2}), (\ref{appendixsonia4}) and
(\ref{appendixsonia6}),
\[ x^{\delta} \partial_{\tmmathbf{t}}^{\alpha} \partial_x^{\beta} \psi
   =\mathcal{F} \tilde{f} \]
where
\[ \tilde{f} \in \text{$S_{r - 2 \alpha - | \beta | - | \delta |}^{2 \alpha +
   | \beta |, k + 2 \alpha + | \beta | + | \delta |}$} \]
and
\[ | \tilde{f} |_{2 \alpha + | \beta |, k + 2 \alpha + | \beta | + | \delta |,
   p - \alpha, q - | \beta |, r - 2 \alpha - | \beta | - | \delta |} \leqslant
   c_1^{\alpha} c_2^{| \beta |} c_3^{| \delta |} C\| \varphi \|_{k, r} . \]
Hence, for $\tmmathbf{t} \in] - T_2, T_2 [-\{0\}$ and $x \in
\mathbbm{R}^{\nu}$,
\begin{eqnarray*}
  \ve{1} x^{\delta} \partial_{\tmmathbf{t}}^{\alpha} \partial_x^{\beta}
  \tmmathbf{\psi} \sve{1}{}{} \leqslant &  & C_1 |4 \pi \tmmathbf{t} |^{- \nu
  / 2} (1 + |x|)^{2 \alpha + | \beta |} \int_{\mathbbm{R}^{\nu}} (1 + |y|)^{-
  \nu - 1} d y \times \| \varphi \|_{k, r}\\
  \leqslant &  & C_2 | \tmmathbf{t} |^{- \nu / 2} (1 + |x|)^{2 \alpha + |
  \beta |} \times \| \varphi \|_{k, r} .
\end{eqnarray*}

\medskip

\tmtextbf{-} Let $p, q, k' \geqslant 0$. Let us choose $k \in \mathbbm{R}$ and
$r \in \mathbbm{N}$ such that
\[ \left\{ \begin{array}{l}
     k + 2 p + q + k' < \um \nu - 1\\
     \\
     r - 2 p - q - k' \geqslant 0
   \end{array} \text{} . \right. \]
Then, if $\alpha \leqslant p$, $| \beta | \leqslant q$ and $| \delta |
\leqslant k'$, (\ref{appendixsonia8}) is satisfied and
\[ \forall \tmmathbf{t} \in] - T_2, T_2 [-\{0\}, \forall x \in
   \mathbbm{R}^{\nu}, \ve{1} x^{\delta} \partial_{\tmmathbf{t}}^{\alpha}
   \partial_x^{\beta} \tmmathbf{\psi} \sve{1}{}{} \leqslant C_3 | \tmmathbf{t}
   |^{- \nu / 2} (1 + |x|)^{2 p + q} \]
where $C_3$ is a positive number. Let $\bar{k} \in \mathbbm{R}$. Then there
exists $C_4 > 0$, such that, for $\tmmathbf{t} \in] - T_2, T_2 [-\{0\}$ and $x
\in \mathbbm{R}^{\nu}$,
\[ \alpha \leqslant p, | \beta | \leqslant q \Rightarrow \ve{1}
   \partial_{\tmmathbf{t}}^{\alpha} \partial_x^{\beta} \tmmathbf{\psi}
   \sve{1}{}{} \leqslant C_3 | \tmmathbf{t} |^{- \nu / 2} (1 + |x|)^{-
   \bar{k}} . \]
If $\kappa > 0$ and $g$ is a smooth function on $] - T_2, T_2 [-\{0\}$
satisfying, for every $N \in \mathbbm{N}$, $\max_{n \leqslant N} |
\partial_{\tmmathbf{t}}^n g ( \tmmathbf{t}) | \leqslant C_N | \tmmathbf{t}
|^{- \kappa}$, then $g$ is smooth on $] - T_2, T_2 [$ and
\[ \sup_{\tmmathbf{t} \in] - T_2, T_2 [, n \leqslant N} |
   \partial_{\tmmathbf{t}}^n g ( \tmmathbf{t}) | \leqslant c C_{N + [\kappa] +
   2} \]
where $c$ only depends on $\kappa$ and $T_2$. Therefore, for every $n \in
\mathbbm{N}$, $\partial_{\tmmathbf{t}}^n \tmmathbf{\psi} ( \tmmathbf{t},
\cdot) \in \mathcal{S}(\mathbbm{R}^{\nu})$ and for every $(k', q) \in
\mathbbm{R} \times \mathbbm{N}$ there exist $(k, r) \in \mathbbm{R} \times
\mathbbm{N}$ and $C > 0$ such that
\[ \sup_{\tmmathbf{t} \in] - T_2, T_2 [} \| \partial_{\tmmathbf{t}}^n
   \tmmathbf{\psi} ( \tmmathbf{t}, \cdot)\|_{k', q} \leqslant C\| \varphi
   \|_{k, r} . \]
i.e. the mapping $\varphi \longmapsto \tmmathbf{\psi}$ is continuous.

\medskip

\tmtextbf{-} Let us consider the assertion on $\tmmathbf{\psi}
|_{\tmmathbf{t} = 0}$. Let $\gamma \in \mathcal{D}(\mathbbm{R}^{\nu})$ be such
that $\gamma (z) = 1$ if $|z| \leqslant 1$. Let $x \in \mathbbm{R}^{\nu}$.
Then $\varphi = \varphi_1 + \varphi_2$ where $\varphi_1 (y) = \gamma (y - x)
\varphi (y)$ and $\varphi_2 (y) = \lp{1} 1 - \gamma (y - x) \rp{1} \varphi
(y)$. Since both functions belong to the Schwartz space, it suffices to check
the claim for the corresponding $\tmmathbf{\psi}_1$ and $\tmmathbf{\psi}_2$.
Since $\varphi_2$ vanishes on a neighbourhood of $x$, one gets
$\tmmathbf{\psi}_2 ( \tmmathbf{t}, x) =\mathcal{O}( \tmmathbf{t}^{\infty})$ by
integrations by parts. Since the support of the function $\varphi_1$ is
compact, $\tmmathbf{\psi}_2 (\cdot, x) |_{\tmmathbf{t} = 0} = \varphi_2 (x)$.
This proves $\tmmathbf{\psi} |_{\tmmathbf{t} = 0} = \varphi$.

\bigskip

\ \ \ \ \ \ \ \ \ \ \ \ \ \ \ \ \ \ \ \ \ \ \ \ \ \ \ \ \ \ \ \ \ \ \ \ \ \
\ \ \ \ \ REFERENCES

\medskip

[Bals] W. Balser, From divergent power series to analytic functions,
Springer-Verlag.

[Bo-Ca-H\"a-Mi] J.F Bony, R. Carles, D. H\"afner, L. Michel, Scattering theory
for the Schr\"odinger equation with repulsive potential, J. Math. Pures Appl.
84 (2005) 509-579.

[Ha4] T. Harg\'e, Borel summation of the small time expansion of the heat
kernel. The scalar potential case (2013).

[Ha5] T. Harg\'e, Borel summation of the heat kernel with a vector potential
case (2013).

[Ha6] T. Harg\'e, A deformation formula for the heat kernel (2013).

[K-S-W] T. Kuna, L. Streit, W. Westerkamp, Feynman integrals for a class of
exponentially growing potentials, J. Math Phys. 39 (1998) p. 4476-4491, arxiv
0303022v1.

[Ma-Ra] B. Malgrange, J.P. Ramis, Fonctions multisommables, Annales de
l'institut Fourier tome 42, $n^{\circ}$ 1-2 (1992) p. 353-368.

[Re-Si] M. Reed, B. Simon, Methods of modern mathematical physics, vol.2.

[Ru] W. Rudin, Real and complex analysis.

[So] A.D. Sokal, \tmtextup{An improvement of Watson's theorem on Borel
summability}, J. Math. Phys \tmtextbf{21(2)} (1980), 261-263.

\bigskip

D\'epartement de Math\'ematiques, Laboratoire AGM (CNRS), Universit\'e de
Cergy-Pontoise, 95000 Cergy-Pontoise, France.

\[  \]

\end{document}